\documentclass{elsart}
\usepackage{amsbsy,amsmath,amssymb}
\begin{document}
\begin{frontmatter}
\title{LARGE-ORDER ASYMPTOTES FOR DYNAMIC MODELS NEAR EQUILIBRIUM}

\author[1]{Juha Honkonen\corauthref{cor1}}
\ead{Juha.Honkonen@helsinki.fi},
\author[2]{M. V. Komarova}
\ead{komarova@paloma.spbu.ru}
and
\author[2]{M. Yu. Nalimov}
\ead{Mikhail.Nalimov@pobox.spbu.ru}
\address[1]{
Theoretical Physics~Division,
Department~of~Physical Sciences, P.O.~Box~64,FIN-00014
University~of~Helsinki, Finland}
\address[2]{
Department of Theoretical Physics, St.~Petersburg
University, Uljanovskaja 1, St.~Petersburg,
Petrodvorets, 198504 Russia}
\corauth[cor1]{corresponding author}

\begin{abstract}
Instanton analysis is applied to model A of critical dynamics. It is shown that the static instanton
of the massless $\phi^{4}$ model determines the large-order asymptotes of the perturbation expansion of
the dynamic model.
\end{abstract}

\end{frontmatter}


\section{Introduction}
\label{sec:intro}

The knowledge of large-order asymptotic behaviour of perturbation
series of static field-theoretic models is important for
resummation of series for critical exponents and scaling functions \cite{Zinn}.
This behaviour has been thoroughly explored with
the aid of instanton analysis and applied to the resummation problem in the prototypical static $\phi^4$
model~\cite{Lipatov,ZinnBr,TMF}, which has been widely used as a model of critical
behaviour in continuous phase transitions of
ferromagnetic type.

However, little is known about large-order asymptotics in dynamic
field theories constructed from Langevin equations with the aid of
the Martin-Siggia-Rose (MSR) formalism~\cite{MSR}. Recently, it was
stated~\cite{Chetkov,Balkovski} that there is no instanton within
the MSR approach in the Kraichnan model -- which has attracted
considerable attention as a model describing intermittency in
turbulent diffusion~\cite{Falkovich01} -- and that the method of
steepest descent has to be used in Lagrangean variables.

In this paper we propose a method to assess large-order
asymptotic behaviour of dynamic models near equilibrium, i.e. with
Gibbsian static limit (we also will restrict ourselves to models
generated by Langevin equations without mode coupling). We will
use the steepest descent method (instanton approach) to find the
large-order behaviour in a representative model. We will discuss
large-order asymptotes of
correlation functions, response functions and critical exponents.
For equal-time correlation functions this
asymptotic behaviour is shown to coincide with that of the static
instanton approach.

The particular model we deal with in this paper is one of the
standard dynamic $\phi^4$-based models: model A in the
classification of Ref.~\cite{Hohenberg77}. In this model critical
exponents are the same as in the static $\phi^4$ model apart from
the dynamic exponent $z$. The large-order asymptotics of the
dynamic exponent have not been analyzed so far. It should be noted
here that the use of Lagrangean variables becomes prohibitively
difficult in this case due to the essential non-linearity of the
problem (the Kraichnan model is linear in the advected scalar
field).

The present article is organized as follows: in Sec. \ref{sec:dynfieldth}
construction of the MSR field theory corresponding to a non-linear
Langevin equation is reviewed with special emphasis on the treatment of
the functional determinant for the steepest descent method. Existence of the
dynamic instanton and relation of the dynamic instanton solution to the
static one is analysed in Sec. \ref{sec:instanton}. The fluctuation determinant
is calculated in Sec. \ref{sec:fluctuationDet}, whereas Sec. \ref{sec:corrresp}
is devoted to a brief analysis of correlation and response functions.
Results of this paper are summarized in Sec. \ref{sec:conclusion}.

\section{ Dynamic field theory}
\label{sec:dynfieldth}

Consider the Langevin equation
\begin{equation}
\label{LanEq}
\frac{\partial \varphi}{\partial t}+\Gamma \frac{\delta
S}{\delta\varphi}=\xi\,,
\end{equation}
where $\xi$ is a Gaussian random field with zero mean and the
correlation function
\[
\langle\xi(t,{\bf x})\xi(t',{\bf
x}')\rangle=D(x-x')=2\Gamma\delta(t-t')\delta({\bf x}-{\bf x}')\,,
\]
where the shorthand notation $x=(t, {\bf x})$ has been used. In equation
(\ref{LanEq}) the action is the static action of arbitrary model
with the known instanton. The most interesting example is the
massless $\varphi^4$ model:
\begin{equation}
\label{staticS}
S=\frac{1}{2}\partial\varphi\partial\varphi+\frac{g}{4!}\varphi^4\,.
\end{equation}
Here and henceforth, all necessary integrals and sums are implied.
We find it convenient to start the analysis on a finite time
interval $[t_0,T]$, but in the full Euclidian space ${\bf R}^D$.
In the MSR approach, functional
integrals for correlation and response functions are calculated
with the ''measure''
\begin{equation}
\label{DDJmeasure}
\mathfrak{M}[\varphi,\varphi']=
\mathcal{D}\varphi \mathcal{D}\varphi ' \det M\, e^{-\overline{S}+A_\varphi \varphi+A_{\varphi'}\varphi'}\,,
\end{equation}
where the dynamic action is of the De Dominicis-Janssen
form~\cite{DeDominicis76}
\begin{equation}
\label{DDJaction}
\overline{S}=-\frac{1}{2}\varphi' D\varphi'
+\varphi'\left(\frac{\partial\varphi}{\partial t} +\Gamma\frac{\delta
S}{\delta\varphi}\right)\,.
\end{equation}
The operator determinant $\det M$ in (\ref{DDJmeasure}) has different interpretations.
Most often it is written as a result of a formal change of variables in a
$\delta$ functional imposing condition (\ref{LanEq}), which yields~\cite{Zinn,DeDominicis76}
\[
M=\frac{\partial\delta(x-x') }{\partial t}+\Gamma \frac{\delta^2
S}{\delta\varphi(x)\delta\varphi(x')}
\]
for the operator $M$, after which the determinant is calculated in a formal loop
expansion as
\begin{equation}
\label{det1/2}
\det M=\det\left[\left(\frac{\partial}{\partial t}+\Gamma K\right)\delta(x-x') \right]
\exp\left[{ 1\over 2}\,\iint\!dxdx'\,\frac{\delta^2
S_I}{\delta\varphi(x)\delta\varphi(x')}\right]
\end{equation}
where $K$ is the differential operator of the free-field part of the static action
[e.g. $K=-\nabla^2$ for action (\ref{staticS})] and $S_I$ its interaction part. An annoying feature of this
straightforward method is that the normalization determinant
$\det\left[(\partial_t+\Gamma K)\delta(x-x')\right]=\det M_0$ is that of a non-self-adjoint
operator. The explicit content of this factor is unimportant in the ordinary perturbation theory,
because in the calculation of all expectation values it simply cancels out in each term
separately. However, in instanton calculus the saddle-point solution which serves as the
expansion point for a steepest-descent calculation of the functional integral is different from that
of the ordinary perturbation expansion. Therefore, the possible cancellation of determinants requires
additional analysis, which is much more convenient to carry out in terms of determinants of
self-adjoint operators.

It is quite possible and completely consistent to arrive at a functional measure, in which the
normalizing determinant is that of a self-adjoint operator. To this end, recall the basics
of the MSR construction, in which classical quantities are replaced by field
operators $\hat{\psi}$, $\hat{\psi}'$ with the usual bosonic commutation relations
\{normalized as $[\hat{\psi}(t,{\bf x}),\hat{\psi}'(t,{\bf x}')]=\delta({\bf x}-{\bf x}')$\}
and the dynamics are given by the operator equations~\cite{MSR}
\[
{\partial \hat{\psi}\over\partial t}=\left[\hat{\psi},\hat{\mathcal{ H}}\right]\,,\qquad
{\partial \hat{\psi}'\over\partial t}=\left[\hat{\psi}',\hat{\mathcal{ H}}\right]\,,
\]
where the MSR "Hamiltonian" is
\begin{equation}
\label{MSRH}
\hat{\mathcal{ H}}=-{1\over 2}\,\hat{\psi}'D\hat{\psi}'+\Gamma\hat{\psi}'
\left[{\delta S(\varphi)\over \delta\varphi}\right]\Biggl\vert_{\varphi\to \hat{\psi}}\,.
\end{equation}
We have chosen the MSR Hamiltonian in the form (\ref{MSRH}), which is slightly different from the originally
proposed (although this possibility was discussed in the original paper \cite{MSR}), because we want
to write the free-field part of the dynamic action in the form of a formally convergent Gaussian integral at
the outset. The chosen MSR Hamiltonian corresponds to the stage, where all integrations over the random
source $\xi$ of Langevin equation (\ref{LanEq}) have been carried out, and -- integrated over time --
it yields the dynamic action (\ref{DDJaction}).

With the use of the traditional operator formalism the generating functional
of Green functions of these fields may be expressed in the functional
form~\cite{Vasilev98}
\begin{equation}
\label{P}
G(A_\varphi,A_{\varphi'})=\exp\left[{1\over 2}{\delta\over\delta\Psi}\Delta{\delta\over\delta\Psi}\right]
\exp\left[\overline{S}_I+A_\varphi\psi+A_{\varphi'}\psi'\right]\bigl\vert_{\Psi=0}\,,
\end{equation}
where $\overline{S}_I$
is the time integral of the interaction part of the MSR Hamiltonian (\ref{MSRH}) in
the interaction picture,
whereas the kernel of the functional differential operator -- written in terms of the
two-component field $\Psi=(\psi,\psi')$ -- is the propagator
\begin{multline}
\label{kernel}
\Delta(x,x')=\left[
\begin{array}{cc}
\Delta_{11}(x,x')& \Delta_{12}(x,x')\\
\Delta_{21}(x,x')& 0
\end{array}
\right]\\
=\left\{
\begin{array}{cc}
\left[\left(\partial_t+\Gamma K\right)^{-1}D
\left(\partial_t+\Gamma K\right)^{-T}\right](x,x')& \left(\partial_t+\Gamma K\right)^{-1}(x,x')\\
\left(\partial_t+\Gamma K\right)^{-T}(x,x')& 0
\end{array}
\right\}\\
=\left\{
\begin{array}{cc}
\left[M_0^{-1} DM_0^{-T}\right](x,x')& M_0^{-1}(x,x')\\
M_0^{-T}(x,x')& 0
\end{array}
\right\}\,.
\end{multline}
In this approach the functional integral arises from expressing the functional differential
operator in (\ref{P}) as the Gaussian functional integral:
\begin{multline}
\label{Gaussian}
\exp\left[{1\over 2}{\delta\over\delta\Psi}\Delta{\delta\over\delta\Psi}\right]=
\sqrt{\det M_0^TM_0}\\
\times\int\!\mathcal{ D}\varphi\mathcal{D}\varphi'\exp\left[-{ 1\over  2}{\Phi \mathcal{ K}\Phi}
+\varphi{\delta\over\delta\psi}+\varphi'{\delta\over\delta\psi'}\right]\,,
\end{multline}
where $\Phi=(\varphi,\varphi')$ and
the operator $\mathcal{K}$ satisfies $\mathcal{K}\Delta=1$ with the appropriate boundary
conditions. This is the only place where a functional determinant arises and now it is that
of a self-adjoint operator $M_0^TM_0$ completely determined by the free-field part
of the MSR Hamiltonian.

It might be asked what happened to the field-dependent extra
factor in (\ref{det1/2})? The answer depends on the definition of the
time-ordered product ($T$ product) at coinciding time arguments. If the time-ordered product
at equal times is defined as the symmetric product of the field operators,
then the interaction functional $\overline{S}_I$ in (\ref{P}) is the Sym form~\cite{Vasilev98}
of the interaction part of the MSR Hamiltonian. The Sym form of the MSR Hamiltonian (\ref{MSRH})
contains terms quadratic in fields which exactly correspond to the field-dependent
factor in (\ref{det1/2}). It is quite possible, and here in fact more convenient, to
define the coinciding-time time-ordered product as the normal product~\cite{Vasilev98}
of the field operators. In this case the interaction functional $\overline{S}_I$
is taken in the normal form, which in the perturbation
theory implies the condition that there are no graphs with closed
loops of the retarded propagator $(\partial_t+\Gamma K)^{-1}(x,x')$ attached
to the interaction vertex. This amounts to vanishing retarded free-field propagator at coinciding times,
and no additional factors to the determinant $\det M_0^TM_0$ are generated.
Henceforth, we will use the latter approach which allows to choose the determinant factor
of (\ref{DDJmeasure}) in the field-independent form $\det M=\sqrt{\det M_0^TM_0}$.

In the instanton approach the space of integration
in the functional integral is chosen using the properties of the saddle point and may
well be different from that of the ordinary perturbation theory. Since the starting point
is the functional integral constructed for perturbation theory, let us recall what the
appropriate space of integration is in that case.
In the proper choice of the formal space of integration
there are two main points to be taken into account \cite{Vasilev98}: the operator $\mathcal{K}$
must be nondegenerate and the kernel of the uniquely defined operator $\mathcal{K}^{-1}$ must
be the propagator $\Delta$ used for the construction of the perturbation theory. For the
model generated by the static action (\ref{staticS}) the usual diffusion kernel
in unbounded space
\[
\left(\partial_t-\Gamma \nabla^2\right)^{-1}(x,x')={\theta(t-t')\,
\exp\left[-{\displaystyle({\bf x}-{\bf x}')^2\over \displaystyle 4\Gamma(t-t')}\right]
\over \displaystyle\left[4\pi\Gamma(t-t')\right]^{d/2}}
\]
is the choice to generate the propagator $\Delta$ in (\ref{kernel}). The space of integration
$E(\Delta)$ conforming to these requirements may be constructed as
\begin{equation}
\label{E}
\left(
\begin{array}{c}
\varphi\\
\varphi'
\end{array}
\right)
=\left(
\begin{array}{cc}
\Delta_{11}& \Delta_{12}\\
\Delta_{21}& 0
\end{array}
\right)\,
\left(
\begin{array}{c}
\eta\\
\eta'
\end{array}
\right)
=
\left(
\begin{array}{l}
\Delta_{11}\eta+\Delta_{12}\eta'\\
\Delta_{21}\eta
\end{array}
\right)\,,
\end{equation}
where $\eta$ and $\eta'$ are fields vanishing at the boundaries
(i.e. at $\vert{\bf x}\vert\to\infty$, $t=t_0$ and $t=T$) to ensure
all partial integrations in the differential operators of $\mathcal{K}$ without surface terms.
It is worth noting that $\varphi'(T)=0$,
but both fields have finite (and real-valued) initial values at $t=t_0$.

In terms of fields vanishing at the boundaries the right-hand side of equation (\ref{Gaussian})
assumes a bit different form giving rise to the representation
\begin{multline}
\label{GaussianE}
\exp\left[{1\over 2}{\delta\over\delta\Psi}\Delta{\delta\over\delta\Psi}\right]=
\sqrt{\det \Delta_{12}\Delta_{21}}
\int\!\mathcal{D}\eta'\mathcal{D}\eta\exp\left[-{ 1\over  2}\eta\Delta_{11}\eta
-\eta\Delta_{12}\eta'\right]\\
\times\exp\left[
\left(\Delta_{11}\eta+\Delta_{12}\eta'\right){\delta\over\delta\psi}+\Delta_{21}\eta{\delta\over\delta\psi'}\right]\,,
\end{multline}
in which the Gaussian functional integral is taken over fields vanishing at boundaries. Nevertheless,
formal convergence requires it to be understood as an iterated integral in the order indicated in (\ref{GaussianE}),
the field $\eta$ to be real and the field $\eta'$ purely imaginary.

\section{Instanton analysis}
\label{sec:instanton}

Let us calculate the parametric integral expressing the $N$th
order contribution to perturbation expansion in $g$
of the $k$-point correlation function
\begin{equation}
\label{Norder}
\frac{1}{2\pi i}\oint \frac{dg}{g}
\frac{\displaystyle\iint \!\mathcal{D}\varphi \mathcal{D}\varphi '
\varphi({\bf{x_1}},t_1)\ldots
\varphi({\bf{x_k}},t_k)\, e^{\displaystyle-\overline{S}-N \lg g}}{\displaystyle\iint\!
\mathcal{D}\varphi \mathcal{D}\varphi '\,
e^{\displaystyle-\overline{S}_0}}
\end{equation}
by the method of steepest descent in the variables $g$ and
$\varphi$. We have written the normalizing determinant
in the form of the functional integral
\[
\left(\det M_0^TM_0\right)^{-1/2}= \iint\!\mathcal{D}\varphi \mathcal{D}\varphi '\,
e^{\displaystyle -\overline{S}_0}
\]
with the free-field part of the dynamic action
\[
\overline{S}_0=-\frac{1}{2}\varphi' D\varphi'
+\varphi'\left(\frac{\partial\varphi}{\partial t} +\Gamma K\varphi\right)
\]
in order to make explicit the cancellation of Jacobians accompanying
several forthcoming changes of variables. We will show that the instanton
solution for expression (\ref{Norder}) may be constructed in close relation to the instanton solution
of the corresponding equilibrium model. Therefore, it is worth reminding that the
equilibrium expression for the $k$-point correlation function
without vacuum loops [there are no vacuum loops in the dynamic correlation function
(\ref{Norder})] of the field $\varphi$ is recovered in the limit $t_0\to -\infty$
for $t_1=t_2=\ldots =t_k$.

Applying the instanton approach we make -- in analogy with the usual instanton analysis of
the $\varphi^4$ model -- the change of variables $\varphi\rightarrow
\sqrt{N}{\varphi}$, $\varphi'\rightarrow
\sqrt{N}{\varphi}'$, $g\rightarrow g/N$; in case of any other model it
is convenient to scale the field variables extracting the
dependence of $N$ in the expression for $\overline{S}$. The
corresponding Jacobians from the numerator and the denominator of
(\ref{Norder}) cancel. The stationarity equations for the
exponential of the numerator are
\begin{align}
\label{statEqPhi}
\frac{\delta\overline{S}}{\delta\varphi}&=
-{\partial \varphi '\over\partial t}+\Gamma\frac{\delta^2 S}{\delta \varphi^2}\varphi '=0,\\
\label{statEqPhiPrime} \frac{\delta\overline{S}}{\delta\varphi'}&=
-D\varphi'+{\partial
\varphi\over \partial t}+\Gamma \frac{\delta S}{\delta\varphi}=0\,,\\
\label{statEqG}
{\partial \overline{S}\over \partial g}&=-\frac{1}{g}\,.
\end{align}
To construct the instanton solution, consider the following auxiliary
equation [note the change of sign of the time derivative in comparison with the the Langevin equation
(\ref{LanEq}) and the corresponding stationarity equation (\ref{statEqPhiPrime})]
\begin{equation}
\label{stat1.2}
-{\partial \varphi\over \partial t} +\Gamma \frac{\delta S}{\delta \varphi}=0
\end{equation}
and assume for the time being that there is a
solution $\varphi_d$ conforming the Cauchy condition
for the {\em final} time instant
$\varphi_d(T,{\bf x})=\varphi_f({\bf x})$ with a sufficiently smooth function
$\varphi_f({\bf x})$.
Substitution of this solution to equation (\ref{statEqPhiPrime}) yields
\[
-D\varphi'+{\partial
\varphi_d\over \partial t}+\Gamma \frac{\delta S}{\delta\varphi}\Biggl|_{\varphi\to\varphi_d}=
-D\varphi'+2{\partial
\varphi_d\over \partial t}=0
\]
which leads to the nontrivial $\varphi'_d$
\begin{equation}
\label{dPrime}
\varphi_d'=2D^{-1}{\partial
\varphi_d\over \partial t}\,.
\end{equation}
On the other hand, equation (\ref{statEqPhi})
on this solution is satisfied automatically: indeed, taking the time derivative of equation (\ref{stat1.2}) yields
\[
{\partial\over \partial t}{\partial \varphi_d\over \partial t}
-\Gamma \frac{\delta^2 S}{\delta\varphi\delta \varphi}\Biggl|_{\varphi\to\varphi_d}
\negthickspace\negthickspace\negthickspace{\partial \varphi_d\over \partial t}
=\left({\partial\over \partial t} -\Gamma \frac{\delta^2 S}{\delta\varphi\delta
\varphi}\Biggl|_{\varphi\to\varphi_d}\right)
{\partial \varphi_d\over \partial t}=0
\]
which, apart from a multiplicative factor, is equation (\ref{statEqPhi}) on the solution (\ref{dPrime}).

Finally, equation (\ref{statEqG}) imposes an asymptotic condition on $\varphi_d$: it may be
cast in the form (we remind that on the left-hand side of this equation integrations over
time and space in the action are implied and intact)
\begin{multline}
\nonumber
{\partial \overline{S}\over \partial g}=\Gamma\varphi_d'{\partial\over \partial g}\frac{\delta S}{\delta\varphi}
\Biggl|_{\varphi\to\varphi_d}=\int\limits_{t_0}^T\!dt{d\over dt}{\partial S\over \partial g}\Biggl|_{\varphi\to\varphi_d}\\
=\lim\limits_{t\to T}{\partial S(\varphi_d)\over \partial g}
-\lim\limits_{t\to t_0}{\partial S(\varphi_d)\over \partial g}=-\frac{1}{g}
\end{multline}
since $\Gamma \varphi_d'=2\Gamma D^{-1}\partial_t\varphi_d=\partial_t\varphi_d$.
Taking into account the boundary conditions imposed on the solution of equation (\ref{stat1.2}) we immediately
see that on the solution (\ref{stat1.2}) and (\ref{dPrime}) the third saddle-point equation (\ref{statEqG})
reduces to the following  equation for the limiting values of the dynamic instanton
\[
{\partial S(\varphi_f)\over \partial g}-{\partial S(\varphi_0)\over \partial g}=-\frac{1}{g}\,,
\]
of which $\varphi_f$ is the condition for the Cauchy problem and $\varphi_0({\bf x})=\varphi_d(t_0,{\bf x})$.
Since equation (\ref{stat1.2}) gives rise to iterative solution with the advanced diffusion kernel,
we see that
sufficiently small ''initial'' field $\varphi_0$ vanishes in the limit $t_0\to -\infty$ and we thus
recover the usual equation for the static instanton
\begin{equation}
\label{staticG}
{\partial S(\varphi_f)\over \partial g}=-\frac{1}{g}\,,\qquad t_0\to -\infty\,,
\end{equation}
which allows to identify the so far unspecified Cauchy value of the dynamic instanton
with the static one:
$\varphi_d(T,{\bf x})=\varphi_f({\bf x})=\varphi_{st}({\bf x})$.

Substitution of the solution $\varphi_d$ and $\varphi'_d$ determined by equations(\ref{stat1.2}), (\ref{dPrime})
and (\ref{staticG}) in dynamic action (\ref{DDJaction}) leads to the remarkable result that
the dynamic action on the dynamic instanton solution asymptotically coincides with the static action
on the static instanton:
$$
\overline{S}(\varphi_d,\varphi_d')=S(\varphi_{st})-S(\varphi_0)
\xrightarrow[t_0 \to -\infty]{\hbox{}}S(\varphi_{st})\,.
$$
Thus, we arrive at the conclusion that the exponential factor in our steepest descent
analysis of correlation function (\ref{Norder}) as well as pre-exponential factor
asymptotically are the same as in the corresponding equilibrium static theory, the difference
being generated by the (fixed, no averaging assumed) initial condition for the field $\varphi_0$ only.

Let us explain the choice of the auxiliary equation
(\ref{stat1.2}). To consider the asymptotic behaviour of the
instanton $\varphi_d$ at $t\to \pm \infty $ we have to study
dynamic properties of this equation by investigation of its
fixed points. Apart from the assumed solution $\varphi_d$
discussed so far, equation (\ref{stat1.2}) has two rather obvious
time-independent solutions: the trivial solution $\varphi = 0$ and
the static instanton $\varphi_{st}$ which obeys
$\partial_t\varphi_{st}=0$ and $(\delta
S/\delta\varphi)_{\varphi\to\varphi_{st}}=0$. The trivial solution
does not conform to the non-vanishing initial condition, but gives
rise to $\Gamma\delta^2S/\delta\varphi^2=\Gamma K$. This is a
positive definite operator which renders the trivial solution
stable in the stationary limit $t_0\to -\infty$. The second
variation of the dynamic action for the stationary instanton is
$\Gamma \left(K-{\displaystyle |g|\over\displaystyle
2}\varphi^2_{st}\right)$. This operator is well known from the
instanton analysis of the usual $\varphi ^4$ model~\cite{Zinn}. It
has at least one negative eigenvalue which determines the
direction along which the solution approaches the stationary
instanton at $t\to\infty $. Such a situation with the
time-independent solutions is a heavy argument in favor of the
existence of the dynamic instanton which behaves as
\[
\lim\limits_{t\to t_0}\varphi _d =\varphi_0\xrightarrow[t_0 \to -\infty]{\hbox{}}0\,, \qquad
\lim\limits_{t\to \infty}\varphi _d =\varphi_{st}\,,
\]
even when the explicit solution is not known.

As an example of this statement, consider a simple
dynamic model, viz. the zero-dimensional $\varphi^4$ model.
The dynamic action corresponding to Eq.
(\ref{LanEq}) for the "field" $\varphi(t)$ (without ${\bf x}$
dependence) is  $S={\displaystyle \tau\over\displaystyle 2} \varphi^2+{\displaystyle g\over\displaystyle 4!}\varphi^4$.
The equilibrium limit of such a model is
the famous one-dimensional integral
$$\int\limits_{-\infty}^\infty\! dx \exp\left(-
{\displaystyle \tau\over\displaystyle 2}\,x^2-{\displaystyle g\over\displaystyle 4!}\,x^4\right)$$
which is often considered as an example of
applicability of the
instanton approach to the large-order asymptotic
investigations \cite{Zinn,Lipatov}.  Then equation (\ref{stat1.2})
assumes the form
\[
\partial_t \varphi-\Gamma
\left(\tau\varphi+{g\over 6}\varphi^3\right)=0 
\]
with the solution
\begin{equation}
\varphi_d=\varphi_0\sqrt{\frac{6\tau}{(6\tau+g\varphi_0^2)\exp{[-2\Gamma\tau(t-t_0)]}-g\varphi_0^2}}\,.
\label{model2}
\end{equation}
This dynamic instanton tends to $\varphi_{st}=\sqrt{6\tau/(-g)}$
at $t\to\infty$ and vanishes at $t\to -\infty$.

As to the usual $ \varphi ^{4} $ model in $ 4-\epsilon $
dimension, we have not been able to find a closed-form
representation for the dynamic instanton, but a convergent
iterative solution of equation (\ref{stat1.2}) may be constructed as
follows. As the zeroth-order approximation, take
\[
\varphi_d^{(0)}(t,{\bf x})=\int\!d {\bf x}'{
\exp\left[-{\displaystyle({\bf x}-{\bf x}')^2\over \displaystyle 4\Gamma(T-t)}\right]
\over \displaystyle\left[4\pi\Gamma(T-t)\right]^{d/2}}\,\varphi_{st}({\bf x}')\,,
\]
which obviously is a solution of the linearized equation with the final condition
$\varphi_d^{(0)}(T,{\bf x})=\varphi_{st}({\bf x})$.
Rewriting equation (\ref{stat1.2}) as the integral equation
\[
\varphi_d(t,{\bf x})=-{\Gamma g\over 6}\int\limits_{t}^T\!dt'\int\!d {\bf x}'{
\exp\left[-{\displaystyle({\bf x}-{\bf x}')^2\over \displaystyle 4\Gamma(t'-t)}\right]
\over \displaystyle\left[4\pi\Gamma(t'-t)\right]^{d/2}}\,\varphi_d^3(t',{\bf x}')+
\varphi_d^{(0)}(t,{\bf x})
\]
we immediately see that the next-to-leading term of the iterative sequence is
\[
\varphi_d^{(1)}(t,{\bf x})=-{\Gamma g\over 6}\int\limits_{t}^T\!dt'\int\!d {\bf x}'{
\exp\left[-{\displaystyle({\bf x}-{\bf x}')^2\over \displaystyle 4\Gamma(t'-t)}\right]
\over \displaystyle\left[4\pi\Gamma(t'-t)\right]^{d/2}}\,\left[\varphi_d^{(0)}(t',{\bf x}')\right]^3
\]
with the vanishing final value: $\varphi_d^{(1)}(T,{\bf x})=0$.
Proceeding in the same fashion we arrive at the usual tree-graph
solution of the nonlinear equation (\ref{stat1.2}) with the given
final Cauchy condition instead of the initial condition.

Similar solutions were used in \cite{Suslov,TMF1} for the
massive $ \varphi ^{4}$ model in dimension $4-\epsilon$ in the static limit. There
the tree-graph expansion was sufficient to determine the
asymptotic behaviour of the instanton at both large and small
values of the coordinate argument $\vert{\bf x}\vert$, which were of practical interest.

It should also be noted that -- unlike the usual field-theoretic perturbation expansion --
the number of the graphs at
large order of the tree expansion has no factorial growth. Therefore, this
expansion is convergent in a disk of finite radius and can be
analytically continued. For illustration of this statement,
consider the expansion in the parameter $ g $
of the exact solution of the toy model
(\ref{model2}), which obviously has a finite
radius of convergence.

Usually we are interested in the logarithmic theory. In this case the
model is translation invariant with respect to ${\bf x}$ and
scale invariant with respect to a synchronous dilatation of ${\bf
x}$ and $t$ (the scaling dimension of $t$ being twice the scaling
dimension of ${\bf x}$). Therefore, the instanton has an
arbitrariness of the form
\begin{equation}
\label{proizvol}
\varphi _d({\bf x},t)= \frac1{y^{D/2-1}}f_d\Bigg (
\frac{{\bf x}-{\bf x}_0}{y},\Gamma\frac{t}{y^2} \Bigg),
\end{equation}
where $D$ is the space dimension, ${\bf x}_0$ and $y$ are
arbitrary parameters. The coefficient $\Gamma$ in (\ref{proizvol})
is introduced to make the second argument of $f_d$ dimensionless.
The translation invariance in space and dilatation invariance
forces to lift the corresponding degeneracy in the instanton
solution. Due to the non-vanishing value $\varphi'_d(T)
=2D^{-1}\partial_t\varphi_d\vert_{t=T}$ of the auxiliary field of
the dynamic instanton solution, a shift of the integration space
from the original perturbative integration space (\ref{E}) with
$\varphi'(T)  =0$ to a space with  $\varphi'(T)  \ne0$ is implied
in the functional integrals of the $k$-point correlation functions
(\ref{Norder}).

In this integration space the degeneracies of the instanton may be removed
with the aid of the same unit decomposition
\begin{multline}
\label{1nov} 1=\int\! d^D{\bf x}_0\int\limits _{-\infty}^{+\infty }\!d\ln
y^2\,\delta\Bigg[-{g\over 24}\,\int\! d{ \bf x}\, \varphi_\infty
^4({\bf x})\ln \left({{\bf x}-{\bf x}_0\over
y}\right)^2\Bigg]
\\
\times
\delta ^D\Bigg [ -{g\over 24}\int\! d{\bf x}\,\varphi_\infty
^4({\bf x})({\bf x}-{\bf x}_0)\Bigg ]
\Bigg[-\frac{g}{24}\,\int\! d{ \bf x}\, \varphi_\infty ^4({\bf x})\Bigg ]^{D+1}
\end{multline}
as in the static instanton theory, which then imposes conditions
on the final value of the integration field: $\varphi(T,{\bf x})=\varphi_\infty({\bf x})$.
\{Relation (\ref{1nov}) is written for a scalar field $\varphi $, in a vector case one more
contribution due to the rotational invariance is necessary
\cite{Zinn,Suslov}\}. The
$\delta$ functions introduced by the unit decomposition
(\ref{1nov}) fix the arbitrariness of the instanton completely.

An appropriate change of variables to avoid the spatial translation
invariance and dilatation invariance problem is well known~\cite{Zinn,Lipatov,Suslov}.
Therefore, we will henceforth consider only
specific dynamic properties of the model. All needed operations
to deal with the spatial translation invariance and dilatation invariance problems,
together with spatial arguments of
of fields and spatial integration, are implied and omitted in the following.

Thus, we arrive at the correlation function
\[
\frac{1}{2\pi i}\oint \frac{dg}{g}
\frac{\displaystyle\iint \!\mathcal{D}\varphi \mathcal{D}\varphi '
\varphi({\bf{x_1}},t_1)\ldots
\varphi({\bf{x_k}},t_k)\,{\rm I}\, e^{\displaystyle-\overline{S}-N \lg g}}{\displaystyle\iint\!
\mathcal{D}\varphi \mathcal{D}\varphi '\,
e^{\displaystyle-\overline{S}_0}}\,,
\]
where ${\rm I}$ stands for all
contributions from the unit decomposition (\ref{1nov}).
At the leading order in $N$ we replace all $\varphi({\bf{x}_i},t_i)$ by
$\varphi_d$ in the  pre-exponential factor. In the stationary limit $t_0\to-\infty$ the
initial value of the instanton $\varphi_d(t_0) $
vanishes and in the exponential the static instanton action is recovered at
leading order in $N$.

\section{Calculation of the fluctuation determinant}
\label{sec:fluctuationDet}

During the calculation of the fluctuation integral the usual stretching
of variables $\delta\varphi \rightarrow \delta\varphi/\sqrt{N}$,
$\delta\varphi' \rightarrow \delta\varphi'/\sqrt{N}$
in the fluctuation integration in the numerator is accompanied by a
similar transformation $\varphi \rightarrow\varphi/\sqrt{N}$,
$\varphi' \rightarrow\varphi'/\sqrt{N}$ in
the denominator: the Jacobians thus cancel. We shall not write the
factor due to fluctuations of $g$  explicitly, because after
the standard change of variables in the original dynamic action
$\varphi \rightarrow\varphi/\sqrt{-g}$,
$\varphi' \rightarrow\varphi'/\sqrt{-g}$
fluctuation contributions due to fields $\delta\varphi$ and $\delta\varphi'$ on one hand
and due to the coupling constant $\delta g$ on the other factorize, the latter giving rise to a trivial
numerical factor.

The fluctuation integral over $\delta\varphi$ and $\delta\varphi'$ may be written as
\begin{multline}
\delta \Sigma=
\left\{\displaystyle\iint
\!\mathcal{D}\delta\varphi\mathcal{D}\delta\varphi'
\exp\left[\frac{\displaystyle 1}{\displaystyle 2}\delta\varphi' D\delta\varphi'
-\delta\varphi'\left(\frac{\displaystyle\partial\delta\varphi}{\displaystyle \partial t} +\Gamma K\delta\varphi\right)\right]
\right\}^{-1}\\
\times
\displaystyle\iint\!\mathcal{D}\delta\varphi\mathcal{D}\delta\varphi'\, {\rm I} \exp
\left[\frac{\displaystyle 1}{\displaystyle 2}\delta\varphi'D\delta\varphi'
-\frac{\displaystyle 1}{\displaystyle 2} \delta\varphi\Gamma\varphi'_d\frac{\displaystyle\delta^3 S}{\displaystyle\delta\varphi^3}\bigg
|_{\varphi_{d}}\!\!\! \delta\varphi\right.\\
\left.-\delta\varphi'\left({\partial\over\partial t}+\Gamma \frac{\displaystyle\delta^2
S}{\displaystyle\delta\varphi^2}\right)\bigg |_{\varphi_{d}}\!\!\! \delta\varphi \right]\,.
\label{variation}
\end{multline}
For economy of notation, we have used the same symbols for the integration variables
$\delta\varphi'$ and $\delta\varphi$ in both the denominator and the numerator of (\ref{variation}).
On the dynamic instanton we immediately see that
\[
\frac{\displaystyle\delta^3 S}{\displaystyle\delta\varphi^3}\bigg |_{\varphi_{d}}\!\!\!\Gamma\varphi'_d
=\frac{\displaystyle\delta^3 S}{\displaystyle\delta\varphi^3}\bigg |_{\varphi_{d}}\!\!\!{\partial\varphi_d\over \partial t}
={\partial\over \partial t}\frac{\displaystyle\delta^2 S}{\displaystyle\delta\varphi^2}\bigg |_{\varphi_{d}}\,.
\]
With the aid of this relation and the change of variables
\begin{equation}
\label{varchange}
\delta\varphi'=\psi'+2D^{-1}{\partial \delta\varphi\over \partial t}
\end{equation}
the quadratic form of the exponential of the numerator of fluctuation integral (\ref{variation})
may be expressed as
\begin{multline}
\label{d2S}
-{1\over 2}\,\delta^2_{\varphi,\varphi'}\overline{S}=
\frac{\displaystyle 1}{\displaystyle 2}\,\psi' D\psi'
-\psi'\left(-{\partial\over\partial t}+\Gamma K+\Gamma \frac{\displaystyle\delta^2
S_I}{\displaystyle\delta\varphi^2}\right)\bigg |_{\varphi_{d}}\!\!\! \delta\varphi\\
-\frac{\displaystyle 1}{\displaystyle 2}\, \delta\varphi_{st}\frac{\displaystyle\delta^2 S}{\displaystyle\delta\varphi^2}\bigg
|_{\varphi_{st}}\!\!\! \delta\varphi_{st}
+\frac{\displaystyle 1}{\displaystyle 2}\, \delta\varphi_0\frac{\displaystyle\delta^2 S}{\displaystyle\delta\varphi^2}\bigg
|_{\varphi_0}\!\!\! \delta\varphi_0\,,
\end{multline}
where $\delta\varphi_{st}=\delta\varphi\vert_{t=T}$ and
$\delta\varphi_0=\delta\varphi\vert_{t=t_0}$.
Note the change of sign of the time derivative in the
second term of (\ref{d2S}). We also remind that the term
$\frac{\displaystyle 1}{\displaystyle 2}\, \delta\varphi_{st}\frac{\displaystyle\delta^2 S}{\displaystyle\delta\varphi^2}\Big
|_{\varphi_{st}}\!\!\! \delta\varphi_{st}$ in (\ref{d2S}) describing the effect of asymptotic
(equilibrium) fluctuations would be absent in the
original perturbative integration space.

We intend to calculate the fluctuation integral over the dynamic fluctuations
regarding the term $\psi'\Gamma \frac{\displaystyle\delta^2
S_I}{\displaystyle\delta\varphi^2}\Big |_{\varphi_{d}}\!\!\! \delta\varphi$,
where $S_I$ is the interaction part of the static action, as perturbation. At leading order in
$N$, when only the fluctuation determinant is needed without any other fluctuation contributions,
this trick leads to a sum of closed loops
of advanced diffusion kernels which essentially reduces the purely dynamic fluctuation determinant
to a constant regardless of the properties
of the full differential operator in (\ref{d2S}). The account of dynamic fluctuations becomes highly non-trivial,
of course, when pre-exponential fluctuation terms are included.

To arrive at a tractable perturbative expansion of the dynamic fluctuation determinant
we write the
normalization factor in terms of the same perturbation expansion as indicated in (\ref{d2S}).
Change of variables (\ref{varchange}) casts
the free-field action in the denominator in the form
\begin{equation}
\label{d2S0}
-{1\over 2}\,\delta^2_{\varphi,\varphi'}\overline{S}_0=
\frac{\displaystyle 1}{\displaystyle 2}\,\psi' D\psi'
-\psi'\left(-{\partial\over\partial t}+\Gamma K\right)\delta\varphi
-\frac{\displaystyle 1}{\displaystyle 2}\, \delta\varphi_{st} K\delta\varphi_{st}
+\frac{\displaystyle 1}{\displaystyle 2}\, \delta\varphi_0 K\delta\varphi_0
\end{equation}
with the same free-field part for the perturbative calculation as in (\ref{d2S}).
A shift of the integration space similar to that in the numerator of
the fluctuation integral (\ref{variation}) is implied in (\ref{d2S0}) as well.

The integration space corresponding to perturbative expansion with
(\ref{d2S0}) as the free-field action includes fields constructed with the aid
of the advanced diffusion kernel by the standard prescription, i.e.
\begin{equation}
\label{Eadv}
\left(
\begin{array}{c}
\overline{\delta\varphi}\\
\psi'
\end{array}
\right)
=\left(
\begin{array}{cc}
\Delta_{11}& \Delta_{21}\\
\Delta_{12}& 0
\end{array}
\right)\,
\left(
\begin{array}{c}
\eta\\
\eta'
\end{array}
\right)
=
\left(
\begin{array}{l}
\Delta_{11}\eta+\Delta_{21}\eta'\\
\Delta_{12}\eta
\end{array}
\right)\,,
\end{equation}
where $\eta$ and $\eta'$ are fields vanishing at the boundaries
(i.e. at $\vert{\bf x}\vert\to\infty$, $t=T$ and $t=t_0$).
However,
the propagator structure leads to final values of the fields
$\overline{\delta\varphi}$ and $\psi'$, which become arbitrarily
small, when $T$ grows.
Since in the integration space around the instanton the long-time
asymptotics of the fluctuation field $\delta\varphi\to \delta\varphi_{st}$ is finite
and should eventually be independent of the choice of $T$,
an additional contribution is required
to account for the asymptotic fluctuations described by $\delta\varphi_T$.
The spatial convolution of the advanced diffusion kernel with the
difference of the desired asymptotic field $\delta\varphi_{st}$ and the final value
of $\overline{\delta\varphi}$ will do:
\begin{equation}
\label{varphiT}
\delta\varphi_T(t,{\bf x})=\int\!d {\bf x}'{
\exp\left[-{\displaystyle({\bf x}-{\bf x}')^2\over \displaystyle 4\Gamma(T-t)}\right]
\over \displaystyle\left[4\pi\Gamma(T-t)\right]^{d/2}}\,\left[\delta\varphi_{st}({\bf x}')-
\overline{\delta\varphi}(T,{\bf x}')\right]\,,
\end{equation}
since it is the solution of the homogeneous equation
$$
\left[-{\partial\over \partial t}+\Gamma \left(-\nabla^2\right)\right]\delta\varphi_T(t,{\bf x})=0
$$
with the final value $\delta\varphi_T(T,{\bf x})=\delta\varphi_{st}({\bf x})-\overline{\delta\varphi}(T,{\bf x})$.

Due to this, on the finite interval  $[t_0,T]$ the free-field action on the integration field composed as a sum
of (\ref{Eadv}) and (\ref{varphiT}), i.e.
\begin{equation}
\label{intvar}
\delta\varphi=\overline{\delta\varphi}+\delta\varphi_T=
\Delta_{11}\eta+\Delta_{21}\eta'+\delta\varphi_T
\end{equation}
amounts to
\begin{multline}
\label{d2S0T}
-{1\over 2}\,\delta^2_{\varphi,\varphi'}\overline{S}_0=
\frac{\displaystyle 1}{\displaystyle 2}\,\psi' D\psi'
-\psi'\left(-{\partial\over\partial t}+\Gamma K\right)\overline{\delta\varphi}\\
-\frac{\displaystyle 1}{\displaystyle 2}\, \delta\varphi_T(T) K\delta\varphi_T(T)
+\frac{\displaystyle 1}{\displaystyle 2}\, [\overline{\delta\varphi}_0+\delta\varphi_T(t_0)]
K[\overline{\delta\varphi}_0+\delta\varphi_T(t_0)]\,.
\end{multline}
Here, however, $\lim\limits_{T\to\infty}\delta\varphi_T(T)=\delta\varphi_{st}$ by construction (\ref{varphiT}).
Therefore in (\ref{d2S0T}) the explicit dependence
(apart from the upper limit of all time integrals) on $T$ remains only in the
initial value of the field $\delta\varphi_T(t_0)$, which is finite.
Moreover, from (\ref{varphiT}) it also follows that
$\delta\varphi_T(t_0)\to 0$, when $T\to\infty$. Thus, we may write
the original quadratic form on the ray $[t_0,\infty)$ as
\begin{multline}
\nonumber
-{1\over 2}\,\delta^2_{\varphi,\varphi'}\overline{S}_0=
\lim\limits_{T\to\infty}\Biggl\{
\frac{\displaystyle 1}{\displaystyle 2}\,\psi' D\psi'
-\psi'\left(-{\partial\over\partial t}+\Gamma K\right)\overline{\delta\varphi}\\
-\frac{\displaystyle 1}{\displaystyle 2}\, \delta\varphi_{st} K\delta\varphi_{st}
+\frac{\displaystyle 1}{\displaystyle 2}\, \overline{\delta\varphi}_0
K\overline{\delta\varphi}_0\Biggr\}
\end{multline}
-- an expression which in the functional integral of
the denominator of (\ref{variation}) allows to factorize the asymptotic fluctuations described
by $\delta\varphi_{st}$ from the ''purely dynamic'' fluctuations described by
$\overline{\delta\varphi}$ and $\psi'$ from (\ref{Eadv}).

A similar representation in the
quadratic form of fluctuations (\ref{d2S}) leaves explicit dependence on $\delta\varphi_T$ also
in the ''interaction'' term:
\begin{multline}
\nonumber
-{1\over 2}\,\delta^2_{\varphi,\varphi'}\overline{S}=\lim\limits_{T\to\infty}\Biggl\{
\frac{\displaystyle 1}{\displaystyle 2}\,\psi' D\psi'
-\psi'\left(-{\partial\over\partial t}+\Gamma K\right)\overline{\delta\varphi}
-\psi'\left(
\Gamma \frac{\displaystyle\delta^2
S_I}{\displaystyle\delta\varphi^2}\right)\bigg |_{\varphi_{d}}\!\!\! \left(\overline{\delta\varphi}+\delta\varphi_T\right)\\
-\frac{\displaystyle 1}{\displaystyle 2}\, \delta\varphi_T(T)\frac{\displaystyle\delta^2 S}{\displaystyle\delta\varphi^2}\bigg
|_{\varphi_{st}}\!\!\! \delta\varphi_T(T)
+\frac{\displaystyle 1}{\displaystyle 2}\, \delta\varphi_T(t_0)\frac{\displaystyle\delta^2 S}{\displaystyle\delta\varphi^2}\bigg
|_{\varphi_{0}}\!\!\! \delta\varphi_T(t_0)\Biggr\}\,.
\end{multline}
Again $\lim\limits_{T\to\infty}\delta\varphi_T(T)=\delta\varphi_{st}$ and $\displaystyle\lim\limits_{T\to\infty}\varphi_T(t_0)=0$
by construction. Moreover, as seen from (\ref{varphiT}), the latter limit applies for almost all $t$ as well:
$\displaystyle\lim\limits_{T\to\infty}\varphi_T(t,{\bf x}))=0$, $t_0\le t<T$.
From this it follows that
$$
\psi'\left(
\Gamma \frac{\displaystyle\delta^2
S_I}{\displaystyle\delta\varphi^2}\right)\bigg |_{\varphi_{d}}\!\!\! \delta\varphi_T
=\int\limits_{t_0}^T\!
\psi'(t)\left(
\Gamma \frac{\displaystyle\delta^2
S_I}{\displaystyle\delta\varphi^2}\right)\bigg |_{\varphi_{d}}\!\!\! \delta\varphi_T(t)dt
\operatornamewithlimits{\longrightarrow}_{T\to\infty}0\,.
$$
Therefore,
we see that representation (\ref{varphiT}) and (\ref{intvar})
leads to factorization of the time-independent
asymptotic fluctuations $\delta\varphi_{st}$ and the dynamic fluctuations
$\overline{\delta\varphi}$ and $\psi'$
in the numerator of (\ref{variation}) as well.

Therefore, the fluctuation integral may be expressed in the explicitly factorized form
\begin{multline}
\label{variationT}
\delta \Sigma=
\left[\displaystyle\int
\!\mathcal{D}\delta\varphi_{st}
\exp\left(-\frac{\displaystyle 1}{\displaystyle 2}\delta\varphi_{st}  K\delta\varphi_{st}\right)
\right]^{-1}\!\!\!
\displaystyle\int
\!\mathcal{D}\delta\varphi_{st}{\rm I}\,
\exp\left(-\frac{\displaystyle 1}{\displaystyle 2}\delta\varphi_{st}\frac{\displaystyle\delta^2 S}{\displaystyle\delta\varphi^2}\bigg
|_{\varphi_{st}}\!\!\! \delta\varphi_{st}\right)
\\
\left\{\displaystyle\iint
\!\mathcal{D}\overline{\delta\varphi}\mathcal{D}\psi'
\exp\left[\frac{\displaystyle 1}{\displaystyle 2}\psi' D\psi'
-\psi'\left(-\frac{\displaystyle\partial\overline{\delta\varphi}}{\displaystyle \partial t} +\Gamma K
\overline{\delta\varphi}\right)+\frac{\displaystyle 1}{\displaystyle 2}\, \overline{\delta\varphi}_0
K\overline{\delta\varphi}_0\right]
\right\}^{-1}\\
\times
\displaystyle\iint\!\mathcal{D}\overline{\delta\varphi}\mathcal{D}\psi'\,  \exp
\left[\frac{\displaystyle 1}{\displaystyle 2}\psi' D\psi'
-\psi'\left(-{\partial\over\partial t}+\Gamma K\right)\overline{\delta\varphi}\right.\\
\left.
-\psi'\left(
\Gamma \frac{\displaystyle\delta^2
S_I}{\displaystyle\delta\varphi^2}\right)\bigg |_{\varphi_{d}}\!\!\! \overline{\delta\varphi}
+\frac{\displaystyle 1}{\displaystyle 2}\,\overline{\delta\varphi}_0\frac{\displaystyle\delta^2 S}{\displaystyle\delta\varphi^2}\bigg
|_{\varphi_{0}}\!\!\! \overline{\delta\varphi}_0
\right]\,,
\end{multline}
with the dynamic fields of the structure (\ref{Eadv}).
On the right-hand-side of (\ref{variationT}) the first line yields the familiar static fluctuation determinant
including the unit decomposition (\ref{1nov}).

A non-trivial dependence on the initial time instant $t_0$ remains in (\ref{variationT})
through the non-zero initial value of the fluctuation field $\overline{\delta\varphi}_0$.
This is an integral of the independent variables $\eta$ and $\eta'$
with vanishing boundary conditions from representation (\ref{Eadv}):
$$
\overline{\delta\varphi}_0({\bf x})=\int\limits_{t_0}^\infty\!dt\int\!d{\bf x'}\left[
\Delta_{11}(t_0,t,{\bf x}-{\bf x}')\eta (t,{\bf x}')+
\Delta_{21}(t_0-t,{\bf x}-{\bf x}')\eta' (t,{\bf x}')\right]
$$
and thus leads to fairly complicated propagator structure of the
perturbative expansion of the dynamic contribution in
(\ref{variationT}). This also means that a functional dependence
on the dynamic instanton remains in the fluctuation determinant
for finite $t_0$.  In the asymptotic limit $t_0\to -\infty$ --
which restores the translation invariance with respect to time --
the dependence on the instanton vanishes, however. This takes
place, because due to the properties of the diffusion kernel, in
the limit $t_0\to -\infty$ the initial value of the fluctuation
field vanishes: $\overline{\delta\varphi}_0\to 0$. Therefore, in
this limit the dynamic part of (\ref{variationT}) may be expressed
in a form similar to (\ref{P}):
\begin{multline}
\label{Pdet}
\left\{\displaystyle\iint
\!\mathcal{D}\overline{\delta\varphi}\mathcal{D}\psi'
\exp\left[\frac{\displaystyle 1}{\displaystyle 2}\psi' D\psi'
-\psi'\left(-\frac{\displaystyle\partial\overline{\delta\varphi}}{\displaystyle \partial t} +\Gamma K
\overline{\delta\varphi}\right)\right]
\right\}^{-1}\\
\times
\displaystyle\iint\!\mathcal{D}\overline{\delta\varphi}\mathcal{D}\psi'\, {\rm I} \exp
\left[\frac{\displaystyle 1}{\displaystyle 2}\psi' D\psi'
-\psi'\left(-{\partial\over\partial t}+\Gamma K\right)\overline{\delta\varphi}
-\psi'\left(
\Gamma \frac{\displaystyle\delta^2
S_I}{\displaystyle\delta\varphi^2}\right)\bigg |_{\varphi_{d}}\!\!\! \overline{\delta\varphi}
\right]\\
=\exp\left[{1\over 2}{\delta\over\delta\Psi}\Delta^T{\delta\over\delta\Psi}\right]
\exp\left[-\psi'\left(
\Gamma \frac{\displaystyle\delta^2
S_I}{\displaystyle\delta\varphi^2}\right)\bigg |_{\varphi_{d}}\!\!\! \psi\right]\Biggl\vert_{\Psi=0}\,.
\end{multline}
Here $\Delta^T$ is obtained from the propagator kernel (\ref{kernel}) by replacing the retarded diffusion kernel
by the retarded one and vice versa. Inspection reveals that
the perturbation expansion of (\ref{Pdet}) contains only closed loops of
the advanced propagator
which all vanish
(including the closed single loop, due to our definition of the $T$ product at coinciding times).
Therefore, we immediately see that perturbatively the expression in (\ref{Pdet}) is equal to unity
and the dynamic fluctuation determinant (\ref{variation}) is equal to the static one in the limit
$t_0\to-\infty$:
\begin{equation}
\label{variationS}
\lim\limits_{t_0\to -\infty}\delta \Sigma=
{\displaystyle\int
\!\mathcal{D}\delta\varphi_{st}{\rm I}\,
\exp\left(-\frac{\displaystyle 1}{\displaystyle 2}\delta\varphi_{st}\frac{\displaystyle\delta^2 S}{\displaystyle\delta\varphi^2}\bigg
|_{\varphi_{st}}\!\!\! \delta\varphi_{st}\right)
\over
\displaystyle\int
\!\mathcal{D}\delta\varphi_{st}
\exp\left(-\frac{\displaystyle 1}{\displaystyle 2}\delta\varphi_{st}  K\delta\varphi_{st}\right)}\,.
\end{equation}
Thus, not only the pre-exponential and exponential factors in the large $N$ asymptote
of (\ref{Norder}) coincide with the static expression, but the fluctuation integrals
with the chosen regularization of determinants coincide as well.

\section{Dynamic part of the correlation function and response function}
\label{sec:corrresp}

We have shown that dynamic correlation functions coincide with the
static ones at leading order in $1/N$ in the limit $t_0\to
-\infty$, when the model is translational invariant in time.
Leading-order contributions to response functions vanish in this
case, however. To see this, consider the $N$th-order contribution
to the response function
\begin{equation}
\label{NorderR}
\frac{1}{2\pi i}\oint \frac{dg}{g}
\frac{\displaystyle\iint \!\mathcal{D}\varphi \mathcal{D}\varphi '
\varphi({\bf{x}}_1,t_1)
\varphi'({\bf{x}}_2,t_2)\,{\rm I}\, e^{\displaystyle-\overline{S}-N \lg g}}{\displaystyle\iint\!
\mathcal{D}\varphi \mathcal{D}\varphi '\,
e^{\displaystyle-\overline{S}_0}}\,.
\end{equation}
On the instanton solution (\ref{stat1.2}) and (\ref{dPrime}) the pre-exponential factor
in (\ref{NorderR}) includes the factor
$$
\varphi_d'({\bf{x}}_2,t_2)=2D^{-1}{\partial
\over \partial t_2}\varphi_d({\bf{x}}_2,t_2)
$$
which vanishes in the limit $t_0\to -\infty$ due to the necessary asymptotic properties of the
instanton, which approaches a finite time-independent limit at long times after the initial time
instant. Thus, the leading contribution to the response function is given by corrections of order
$1/N$.

To capture dynamic effects it is sufficient to consider only
corrections effected by the replacement $\varphi \to \delta \varphi$
and $\varphi' \to \delta \varphi'$. $1/N$
corrections of any other type lead only to the small changes to the
stationary results obtained previously (without any dynamic dependence).

The fluctuation integral for the pair correlation function is
\begin{multline}
\delta \Sigma_{\varphi\varphi}=
\left\{\displaystyle\iint
\!\mathcal{D}\delta\varphi\mathcal{D}\delta\varphi'
\exp\left[\frac{\displaystyle 1}{\displaystyle 2}\delta\varphi' D\delta\varphi'
-\delta\varphi'\left(\frac{\displaystyle\partial\delta\varphi}{\displaystyle \partial t} +\Gamma K\delta\varphi\right)\right]
\right\}^{-1}\\
\times
\displaystyle\iint\!\mathcal{D}\delta\varphi\mathcal{D}\delta\varphi'\,
\delta\varphi({\bf{x}}_1,t_1)
\delta\varphi({\bf{x}}_2,t_2)\,{\rm I} \exp
\left[\frac{\displaystyle 1}{\displaystyle 2}\delta\varphi'D\delta\varphi'
-\frac{\displaystyle 1}{\displaystyle 2} \delta\varphi\Gamma\varphi'_d\frac{\displaystyle\delta^3 S}{\displaystyle\delta\varphi^3}\bigg
|_{\varphi_{d}}\!\!\! \delta\varphi\right.\\
\left.-\delta\varphi'\left({\partial\over\partial t}+\Gamma \frac{\displaystyle\delta^2
S}{\displaystyle\delta\varphi^2}\right)\bigg |_{\varphi_{d}}\!\!\! \delta\varphi \right]\,.\nonumber
\end{multline}
With the use of the change of variables (\ref{varchange}) and the subsequent decoupling of
dynamic and static fluctuations according to (\ref{Eadv}), (\ref{varphiT}) and (\ref{intvar}),
this fluctuation integral gives rise to the sum
\begin{multline}
\delta \Sigma_{\varphi\varphi}=
{\displaystyle\int
\!\mathcal{D}\delta\varphi_{st}
\delta\varphi_{st}({\bf{x}}_1)
\delta\varphi_{st}({\bf{x}}_2)\,{\rm I}
\,
\exp\left(-\frac{\displaystyle 1}{\displaystyle 2}\delta\varphi_{st}\frac{\displaystyle\delta^2 S}{\displaystyle\delta\varphi^2}\bigg
|_{\varphi_{st}}\!\!\! \delta\varphi_{st}\right)
\over
\displaystyle\int
\!\mathcal{D}\delta\varphi_{st}
\exp\left(-\frac{\displaystyle 1}{\displaystyle 2}\delta\varphi_{st}  K\delta\varphi_{st}\right)}\\
+\left[\displaystyle\int
\!\mathcal{D}\delta\varphi_{st}
\exp\left(-\frac{\displaystyle 1}{\displaystyle 2}\delta\varphi_{st}  K\delta\varphi_{st}\right)
\right]^{-1}\!\!\!
\displaystyle\int
\!\mathcal{D}\delta\varphi_{st}{\rm I}\,
\exp\left(-\frac{\displaystyle 1}{\displaystyle 2}\delta\varphi_{st}\frac{\displaystyle\delta^2 S}{\displaystyle\delta\varphi^2}\bigg
|_{\varphi_{st}}\!\!\! \delta\varphi_{st}\right)
\\
\left\{\displaystyle\iint
\!\mathcal{D}\overline{\delta\varphi}\mathcal{D}\psi'
\exp\left[\frac{\displaystyle 1}{\displaystyle 2}\psi' D\psi'
-\psi'\left(-\frac{\displaystyle\partial\overline{\delta\varphi}}{\displaystyle \partial t} +\Gamma K
\overline{\delta\varphi}\right)+\frac{\displaystyle 1}{\displaystyle 2}\, \overline{\delta\varphi}_0
K\overline{\delta\varphi}_0\right]
\right\}^{-1}\\
\times
\displaystyle\iint\!\mathcal{D}\overline{\delta\varphi}\mathcal{D}\psi'\,
\overline{\delta\varphi}({\bf{x}}_1,t_1)
\overline{\delta\varphi}({\bf{x}}_2,t_2)\,
\exp
\left[\frac{\displaystyle 1}{\displaystyle 2}\psi' D\psi'
-\psi'\left(-{\partial\over\partial t}+\Gamma K\right)\overline{\delta\varphi}\right.\\
\left.
-\psi'\left(
\Gamma \frac{\displaystyle\delta^2
S_I}{\displaystyle\delta\varphi^2}\right)\bigg |_{\varphi_{d}}\!\!\! \overline{\delta\varphi}
+\frac{\displaystyle 1}{\displaystyle 2}\,\overline{\delta\varphi}_0\frac{\displaystyle\delta^2 S}{\displaystyle\delta\varphi^2}\bigg
|_{\varphi_{0}}\!\!\! \overline{\delta\varphi}_0
\right]\,,\nonumber
\end{multline}
where the first term is the static fluctuation correlation function and in the second term the dynamic
part -- multiplied by the static fluctuation determinant (\ref{variationS}) -- may be expressed formally as
$$
\Delta_{12}\left[1+\Gamma \frac{\displaystyle\delta^2
S_I}{\displaystyle\delta\varphi^2}\bigg |_{\varphi_{d}}\!\!\!\Delta_{12}\right]^{-1}D
\left[1+\Delta_{21}\Gamma \frac{\displaystyle\delta^2
S_I}{\displaystyle\delta\varphi^2}\bigg |_{\varphi_{d}} \right]^{-1}\Delta_{21}
$$
in the limit $t_0\to -\infty$.

\section{Conclusion}
\label{sec:conclusion}

In this paper first steps have been taken towards large-order
($N\to\infty$) asymptotic analysis of
non-linear stochastic field theories with the aid of the instanton method.
It is shown that this approach can be used in dynamic models as well
as in the static ones.

For the near-equilibrium model A we have shown that in the
temporal translation-invariant case asymptotic properties at
leading order in $N$ are almost completely determined by the
corresponding static instanton solution and fluctuations
thereabout which leads to factorial behaviour of the large-order
contributions similar to that in the static instanton analysis:
viz. the large-order asymptote for an arbitrary quantity $F$ behaves as
\begin{equation}
\label{c1} F^{[N]}= N!Ca^{N}N^{b},
\end{equation}
where $ F^{[N]} $ is the $N\,$th order of the expansion of $F$ in the
parameter $ e $ ($ e $ is the coupling constant $ g $ or the dimensional regularization parameter $ \epsilon
$), $ a $ and $ b $ are constants known from the analysis of the
static theory \cite{Zinn,Lipatov}, $ C $ is an
amplitude factor --  either a constant or a function of coordinate and time arguments depending on the
quantity $F$. In the static $ \varphi ^{4} $ model
the functions $ C $  for universal quantities have been calculated in
Ref. \cite{TMF1}.

In our treatment it has been quite essential that
the dynamic stochastic problem has a Gibbsian limit, which physically is not
totally unexpected due to the fluctuation-dissipation theorem.

Having been interested
mainly in proving the factorial form
of the large-order asymptotes in the dynamic model considered
we have not dwelled on the problem of effects of renormalization.
However, in the static instanton analysis -- according to \cite{Zinn,Lipatov} -- the renormalization
does non affect the solution of the stationarity equation and at
leading order in $1/N$ it contributes to the $C$-type
constants in correlation and response functions only. Details
of the correct description of the renormalization in the instanton
approach in the minimal-subtraction scheme can be found in Ref. \cite{TMF}.
Similar methods are needed for accurate calculation of the constants
in our asymptotic expressions.

As to the calculation of the dynamic critical exponent $ z $, the
renormalization-group equation for, e.g., the renormalized two-point
function $ W_{2}=\langle\varphi \varphi\rangle^{R} $ at the fixed point
$g_{*} $
\begin{equation}
\label{zRG}
\left(k{\partial\over \partial k}-zt{\partial\over \partial t}+\Delta \right)W_{2}=0,
\end{equation}
may be used. In equation (\ref{zRG}),
$ k $ is the wave number and $ \Delta $ is the scaling dimension known
from the static theory. Details of the transformation of the $ g $
expansion considered above into the $ \epsilon $ expansion are described in
Refs. \cite{Zinn,Lipatov}. From the factorial form of large orders in the $ \epsilon $
expansion the asymptotic estimate for the expansion of the exponent $ z $ follows:
\begin{equation}
\label{zN}
z^{[N]}\sim \frac {k\partial _{k}W_{2}^{[N]}+\Delta ^{[0]}W_{2}^{[N]}
+\Delta ^{[N]}W_{2}^{[0]}-z^{[0]}t\partial _{t}W_{2}^{[N]}+X^{[N]}}
{t\partial _{t}W_{2}^{[0]}},
\end{equation}
where $ X^{[N]} $ stands for corrections in $ 1/N $.
Expression (\ref{zN}) demonstrates that the value $ z^{[N]} $ has the same
factorial form (\ref{c1}) as the static critical exponents.

\end{document}